# A Low Complexity Near-Maximum Likelihood MIMO Receiver with Low Resolution Analog-to-Digital Converters

Arkady Molev-Shteiman, Xiao-Feng Qi, Laurence Mailaender

Radio Algorithms Research, Futurewei Technologies, Bridgewater, NJ, USA

*Abstract* – **Based on a new equivalent model of quantizer with noisy input recently presented in [23], we propose a new low complexity receiver that takes into account the nonlinear distortion (NLD) generated by Analog to Digital converter (ADC) with insufficient resolution. The strength of new model is that it presents the NLD as a function of only the desired part of input signal (without noise). Therefore it can easily be used in a variety of NLD mitigation techniques. Here, as an illustration of this, we use a pseudo-ML approach to detect the original QAM modulation based on the equivalent transfer function and exhaustive search. Simulation results for a single user QAM under flat fading show performance equivalent to a true ML receiver, but with much lower computational complexity. The excellent performance of our receiver is an independent validation of the model [23].**

## I. Introduction

Massive MIMO is an emerging technology capable of improving spectral efficiency of wireless communication by orders of magnitude [1]. However a significant increase in base station antennas implies a proportional increase of cost and power consumption. On the other hand, it was shown that massive MIMO may significantly mitigate the impact of imperfections in the hardware implementation [2] meaning that we may use cheaper and lower energy components to implement Massive MIMO. Quantizers (ADCs and DACs) are important elements in the overall cost and energy budget of massive MIMO base station. It is known that cost and power consumption of the quantizer grows exponentially with the number of quantization bits [3]. Therefore algorithms that reduce quantizer resolution have significant practical importance.

Intuitively, the quantizer resolution should be sufficiently high to ensure that the quantizer noise error power is much lower than the power of thermal noise. As the received SNR at each antenna of a massive MIMO array decreases with an increase in antennas, the tolerable quantization error goes up, and quantizer resolution may be decreased to as low as one bit.

Many contributions which consider uplink Massive MIMO receiver with array of low-resolution ADCs have been published. These works analyze performance of Massive MIMO uplink receiver with low resolution ADC from information theoretic point of view [4]-[7] and propose different methods of channel estimation [8]-[13] and data reception [13]-[15] that take into account the limited ADC resolution.

We identify three main methods of dealing with finite resolution ADCs:

- The Additive Quantization Noise Model (AQNM) [8]-[10] represents the ADC output as a sum of ADC input and quantization error which is uncorrelated with ADC output. This model is correct only if the expectation of the ADC input given ADC output is equal to ADC output, which implies special ADCs designed to match the input Probability Density Function (PDF), e.g. Lloyd-Max quantizer. For the common uniform ADCs it is an approximation which is valid only at high ADC resolution
- The Probabilistic method [11]-[13] searches for a desired input signal that maximizes the likelihood of observing the given ADC output vector.
- The Bussgang decomposition method [6] and [7] considers the ADC as a non-linear element. It represents the ADC output as a sum of ADC input scaled by a certain gain and non-linear distortions that are uncorrelated with the input signal.

The new equivalent model [23] is the extension of the Bussgang decomposition for the scenario when the ADC input signal is the sum of the desired signal and the additive noise. In the original Bussgang decomposition, the NLD is a function of the desired signal statistics, the noise statistics and the sum of the desired input signal and the noise. In contrast, the new decomposition [23] represents the NLD as a function of the desired signal statistics, the noise statistics and the only the desired part of input signal.

According to [23], if the ADC resolution is sufficiently large and input SNR sufficiently low, then the non-linear distortion (NLD) that such ADC introduces are negligible and conventional MIMO receiver that does not take into account influence of low resolution ADC may work without significant performance degradation. However if the input SNR exceeds a certain limit, NLD becomes dominant and causes dramatic performance degradation.

The minimum ADC resolution required for NLD to be negligible is obtained through the method of [23]. .

In this paper we show it is possible to operate with resolution below this limit, if the NLD is modeled and accounted for using the equivalent non-linear transfer function developed in [23]. This may include iterative NLD cancelation [16]-[20], or Minimum Mean Square Error (MMSE) rejection [21], or Maximum Likelihood decoding that takes into account received signal non-linearity [22]. We present an example of the ML Receiver approach. It has the approximately the same performance as a direct ML receiver for low resolution ADC presented in [12]. However, because it operates with Maximum Ratio Combining (MRC) results, it has much lower complexity.

## II. SYSTEM MODEL

The Analog to Digital Converter (ADC) performs the quantization operation which is given by expression,

$$Q(s) = \begin{cases} +(R-1)\cdot\Delta/2 & IF\ (s \geq +(R-2)\cdot\Delta/2) \\ -(R-1)\cdot\Delta/2 & IF\ (s < -(R+2)\cdot\Delta/2) \\ \Delta\cdot round((s/\Delta)+0.5)-1 & ELSE \end{cases} \quad (1.1)$$

where $\Delta$ is quantization step, $round(\ )$ denotes rounding operation and $R$ is the total number of possible quantizer outputs which may get values:

$$q_r = (2\cdot r - R - 1)\cdot\Delta/2 \text{ for } (r=1,2,...,R) \quad (1.2)$$

The number of quantizer bits number equals $\log_2(R)$.

We assume without loss of generality that $\Delta = 2$.

The 2 bit quantizer transfer function is shown in Figure 1

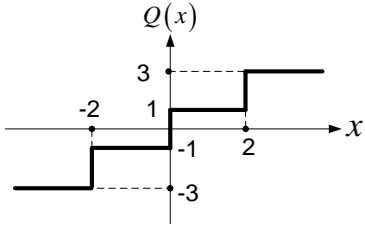

Figure 1 2 bit quantizer transfer function.

Let us define the complex ADC operation as:

$$\tilde{Q}(s) = Q(\text{Re}(s)) + j\cdot Q(\text{Im}(s)) \text{ where: } j = \sqrt{-1} \quad (1.3)$$

We use a superscript '~' to denote a complex quantity.

Let us consider 'all digital' uplink MIMO receiver equipped with $M$ antennas where output of each antenna $m$ complex ADC is given by:

$$\tilde{s}_O(m) = \tilde{Q}(\tilde{s}_I(m) + \tilde{n}_I(m)) \text{ where } \tilde{s}_I(m) = \sum_{k=1}^{K}\tilde{h}_k(m)\cdot\tilde{x}_k \quad (1.4)$$

Where $\tilde{s}_I(m)$ is the desired signal of complex ADC $m$ that we want to pass through ADCs with minimum distortion, $\tilde{n}_I(m)$ is the input noise of complex ADC $m$, $\tilde{x}_k$ is the signal of user $k$, $\tilde{h}_k(m)$ is the channel between user $k$ and antenna $m$ and $K$ is the number of users.

In matrix form we may write expressions (1.4) as:

$$\tilde{S}_O = \tilde{Q}(\tilde{H}\cdot\tilde{X} + \tilde{N}_I) \quad (1.5)$$

Where: $\tilde{X}$ is the input symbols vector with length $K$, $\tilde{H}$ is the channel matrix with size $M \times K$, $\tilde{S}_O$ is the ADC outputs vector with length $M$, $\tilde{N}_I$ is input noise vector with length $M$ and $\tilde{Q}(\ )$ in this context is element-wise complex quantization.

We assume that input noise of each complex ADC is independent circularly symmetric complex Gaussian random variable with zero mean and variance $\sigma_N^2$.

In this paper we consider only the case when input signal $\tilde{x}_k$ is QAM signal. We denote the number of possible QAM symbols (QAM level) as $N_{QAM}$.

## III. MAXIMUM LIKELIHOOD RECEIVERS FOR FINITE RESOLUTION ADC

If the ADC resolution is sufficiently large we may neglect quantization error and use a conventional (naïve) Maximum Likelihood (ML) MIMO receiver given by:

$$\hat{\tilde{X}} = \arg\min_{\tilde{X}}\left((\tilde{S}_O - \tilde{H}\cdot\tilde{X})^H\cdot(\tilde{S}_O - \tilde{H}\cdot\tilde{X})\right) =$$
$$= \arg\min_{\tilde{X}}\left((\tilde{Y} - \hat{\tilde{Y}}(\tilde{X}))^H\cdot\tilde{A}\cdot(\tilde{Y} - \hat{\tilde{Y}}(\tilde{X}))\right) \quad (2.1)$$

where $(\ )^H$ denotes conjugate transpose matrix operation, $\tilde{Y}$ is the Maximum Ratio Combining (MRC) output vector, $\hat{\tilde{Y}}(\tilde{X})$ is the MRC constellation lookup table pre-calculated for all $N_{QAM}^K$ possibilities of vector $\tilde{X}$ and $\tilde{A}$ is the pre-calculated inverse autocorrelation matrix:

$$\tilde{Y} = \tilde{H}^H\cdot\tilde{S}_O \quad (2.2)$$

$$\hat{\tilde{Y}}(\tilde{X}) = \tilde{H}^H\cdot\tilde{H}\cdot\tilde{X} \quad (2.3)$$

$$\tilde{A} = inv(\tilde{H}^H\cdot\tilde{H}) \quad (2.4)$$

The number of complex multiplications required by the naïve ML receiver is equal to:

$$C = C_{MRC} + C_{DIST}\cdot N_{QAM}^K \quad (2.5)$$

where: $C_{MRC} = M\cdot K$ is the MRC calculation complexity.

$C_{DIST} = K^2 + K$ is the distance calculation complexity.

The ML decoder taking into account quantization error chooses such a data vector $X$ that maximizes the likelihood of the observed ADC output

$$\hat{\tilde{X}} = \arg\max_{\tilde{X}}\left(\Pr(\tilde{S}_O | \tilde{S}_I = \tilde{H}\cdot\tilde{X})\right) \quad (2.6)$$

Where:

$$\Pr(\tilde{S}_O | \tilde{S}_I = \tilde{H}\cdot\tilde{X}) = \prod_{m=1}^{M}\Pr\left(\tilde{s}_O(m)\bigg|\tilde{s}_I(m) = \sum_{k=1}^{K}\tilde{h}_k(m)\cdot\tilde{x}_k\right) \quad (2.7)$$

Where:

$$\Pr(\tilde{s}_O|\tilde{s}_I) = \Pr(\text{Re}(\tilde{s}_O)|\text{Re}(\tilde{s}_I))\cdot\Pr(\text{Im}(\tilde{s}_O)|\text{Im}(\tilde{s}_I)) \quad (2.8)$$

The probability of ADC output given ADC input is equal to:

$$\Pr(s_O = q_1 | s_I) = \Pr\left((s_I + n_I) < (q_1 + 1)\right) =$$
$$= 0.5\cdot\left(1 + erf\left((q_1 + 1 - s_I)/\sqrt{2\cdot\sigma_N^2}\right)\right)$$
$$\Pr(s_O = q_R | s_I) = \Pr\left((s_I + n_I) \geq (q_R - 1)\right) = \quad (2.9)$$
$$= 0.5\cdot\left(1 - erf\left((q_R - 1 - s_I)/\sqrt{2\cdot\sigma_N^2}\right)\right)$$

For any other $r = 2,...,R-1$

$$\Pr(s_O = q_r | s_I) = \Pr\left((q_r - 1) \leq (s_I + n_I) < (q_r + 1)\right) =$$
$$= 0.5\cdot\left(erf\left(\frac{q_r + 1 - s_I}{\sqrt{2\cdot\sigma_N^2}}\right) - erf\left(\frac{q_r - 1 - s_I}{\sqrt{2\cdot\sigma_N^2}}\right)\right)$$

Where $erf(\ )$ denotes the error function.

Even if we assume that error function calculation is implemented as a look up table and uses no multiplications, the number of complex multiplications of the resulting ML algorithm (2.6) is equal to:

$$C = M \cdot C_{\text{Pr}} \cdot N_{QAM}{}^K \quad (2.10)$$

Where $C_{\text{Pr}} = K+1$ is the complexity to calculate a single ADC output probability.

This is almost $(M/K)$ times higher than (2.5) which makes implementation of this algorithm impractical.

IV. EQUIVALENT MODEL OF QUANTIZER WITH NOISY INPUT

Let us define expectation of complex ADC pair output $\tilde{s}_O$ given ADC pair input $\tilde{s}_I$ signal as the complex ADC pair equivalent transfer function $\tilde{F}(\tilde{s}_I)$:

$$\tilde{F}(\tilde{s}_I) = E[\tilde{s}_O | \tilde{s}_I] \quad (3.1)$$

Because input noise of real and imaginary part of ADC input noise are independent and have identical Probability Density Function (PDF) $p_N(x)$ we may express ADC pair equivalent transfer function (3.1) as:

$$\tilde{F}(\tilde{s}_I) = E\left[\text{Re}(\tilde{s}_O)|\text{Re}(\tilde{s}_I)\right] + j \cdot E\left[\text{Im}(\tilde{s}_O)|\text{Im}(\tilde{s}_I)\right] = \\ = F(\text{Re}(s_I)) + j \cdot F(\text{Im}(s_I)) \quad (3.2)$$

Where $F(s_I)$ is real ADC equivalent transfer function:

$$F(s_I) = E[s_O|s_I] = \int_{x=-\infty}^{\infty} Q(s_I + x) \cdot p_N(x) \cdot dx = \\ = \sum_{r=1}^{R} q_r \cdot \Pr(s_O = q_r | s_I) \quad (3.3)$$

Where probability of certain ADC output given ADC input $\Pr(s_O = q_r | s_I)$ for Gaussian input noise is equal to (2.9).

Let us define complex quantizer equivalent noise as:

$$\tilde{n}_O(m) = \tilde{s}_O(m) - \tilde{F}(\tilde{s}_I(m)) \quad (3.4)$$

In [23] was proven that quantizer equivalent noise is a white process with zero expectation, zero correlation between noise and input signal, and zero correlation between noises in different array quantizers.

$$E[\tilde{n}_O(m)] = 0 \quad \text{for any } m \quad (3.5)$$
$$E[\tilde{n}_O(m) \cdot \tilde{s}_I(n)] = 0 \quad \text{for any } m \text{ and } n \quad (3.6)$$
$$E[\tilde{n}_O(m) \cdot \tilde{n}_O(n)] = 0 \quad \text{for any } (n \neq m) \quad (3.7)$$

We got equivalent model of complex ADC pair with noisy input that represents quantizer output as sum of quantizer input signal that passes through non-linear element with equivalent transfer function and equivalent additive white noise:

$$\tilde{s}_O(m) = \tilde{Q}(\tilde{s}_I(m) + \tilde{n}_I(m)) = \tilde{F}(\tilde{s}_I(m)) + \tilde{n}_O(m) \quad (3.8)$$

The equivalent block diagram of quantizer with noisy input is shown in Figure **2**.

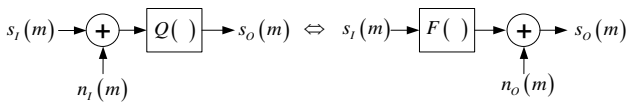

Figure 2 Quantizer equivalent model block diagram

Figure 3 and Figure 4 represent ADC equivalent transfer function calculated according to (3.3) for 1 and 2 bit ADC, respectively, over a range of noise variance. We see that adding noise has linearizing effect on the ADC transfer function.

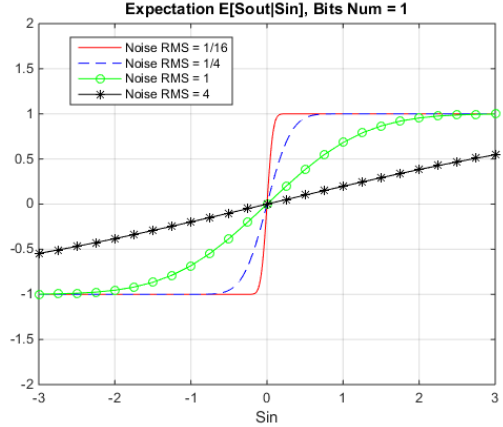

Figure 3  1 bit ADC equivalent transfer function

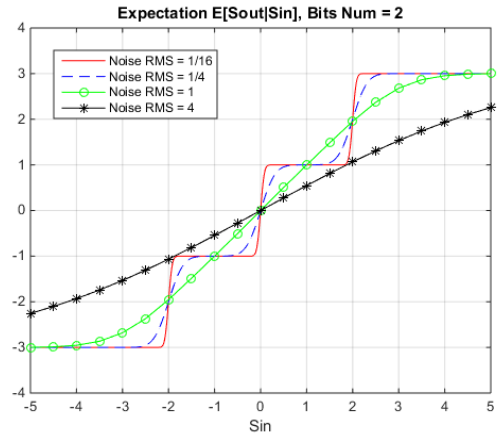

Figure 4  2 bit ADC equivalent transfer function

V. EQUIVALENT ML RECEIVER

According to the proposed equivalent model, output of MRC receiver is equal to:

$$\tilde{Y} = \tilde{H}^H \cdot \tilde{S}_O = \tilde{H}^H \cdot \tilde{F}(\tilde{H} \cdot \tilde{X}) + H^H \cdot \tilde{N}_O \quad (4.1)$$

where $\tilde{N}_O$ is the vector of equivalent additive noise defined by (3.4) with length $M$ and $\tilde{F}(\ )$ is the element-wise equivalent transfer function calculated according to (3.2) and (3.3) Each scalar element of equation (4.1) is given by:

$$\tilde{y}_k = \sum_{m=1}^{M} \tilde{h}_k(m) \cdot \tilde{F}\left(\sum_{k=1}^{K} \tilde{h}_k(m) \cdot \tilde{x}_k\right) + \sum_{m=1}^{M} \tilde{h}_k(m) \cdot \tilde{n}_O(m) \quad (4.2)$$

The equivalent additive noise $\tilde{n}_O(m)$ does not necessarily have Gaussian distribution, however if number of ADCs $M$ is sufficiently large then resulting post MRC additive noise may be approximated as Gaussian. Therefore according to (2.1) an equivalent ML decoder may be approximated as,

$$\hat{\tilde{X}} = \arg\min_{\tilde{X}} \left( \left(\tilde{Y} - \hat{\tilde{Y}}(\tilde{X})\right)^H \cdot \tilde{A} \cdot \left(\tilde{Y} - \hat{\tilde{Y}}(\tilde{X})\right) \right) \quad (4.3)$$

where $\hat{\tilde{Y}}(\tilde{X})$ is a lookup table representing NLD-aware MRC results pre-calculated for all $(N_{QAM})^K$ input possibilities,

$$\hat{\tilde{Y}}(\tilde{X}) = \tilde{H}^T \cdot \tilde{F}(\tilde{H} \cdot \tilde{X}) \tag{4.4}$$

Each element of this table is a vector of NLD-aware MRC outputs predictions, namely $\tilde{F}(\tilde{H} \cdot \tilde{X})$ for a particular $\tilde{X}$. Since this algorithm operate on MRC outputs, its complexity is almost identical to that of the naïve ML receiver(2.5), the only difference being the calculation of the NLD-aware MRC lookup table.

## VI. SIMULATION RESULTS

In order to confirm our equivalent model we provide the following test. We simulate QAM64 uplink communication between a single user and a MIMO base station. The number of receiver antennas and the ADC resolution are simulation parameters. We assume a line of sight channel which is given by:

$$\tilde{h}_1(m) = \exp(j \cdot \pi \cdot \sin(\alpha) \cdot m) \tag{4.5}$$

where the angle of arrival $\alpha$ is a random variable with uniform distribution from $-\pi$ to $+\pi$.

Figure 5 presents 64 MRC output realizations $\tilde{Y}$ obtained from simulation and constellation points predicted by the lookup table (4.4) $\hat{\tilde{Y}}(\tilde{X})$ when angle of arrival $\alpha = \pi/12$, cumulative (sum over all antennas) input SNR is equal to 30dB, MIMO array size is 1024 antennas and ADC resolution is 1 bit. From this figure we may see MRC prediction matches the actual MRC realizations quite well.

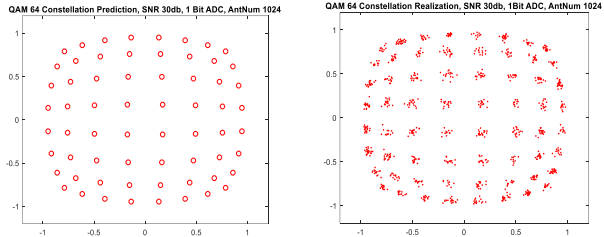

Figure 5  The MRC output constellation points prediction (left) and actual realization (right)

We simulate the following receivers.

- For an array equipped with ideal floating point ADCs (no quantization error) we use conventional (naïve) ML receiver(2.1).
- For an array equipped with low resolution ADCs:
  - The naive ML receiver that ignores the quantization effect (2.1).
  - Brute force ML receiver that incorporates NLD (2.6)
  - NLD-aware equivalent ML receiver (4.3)

Figure 6 and Figure 7 present, respectively, resulting BER as function of cumulative input SNR for an array of 1024 antennas each equipped with a pair of 1-bit ADCs, and for an array of 32 antennas each equipped with a pair of 3-bit ADCs. From these figures we can make the following conclusions:

- For low input SNR, non-linear distortion are small enough and conventional MRC receiver that ignores ADC nonlinearity works quite well. This effect was explained in our previous paper [23].
- For moderate to high SNR, NLD-aware equivalent ML receiver significantly extends the SNR range of reasonable performance
- There is slight performance degradation of the NLD-aware equivalent ML receiver (4.3) relative to the brute force ML receiver (2.6) due to the Gaussian approximation of the post-MRC additive noise.

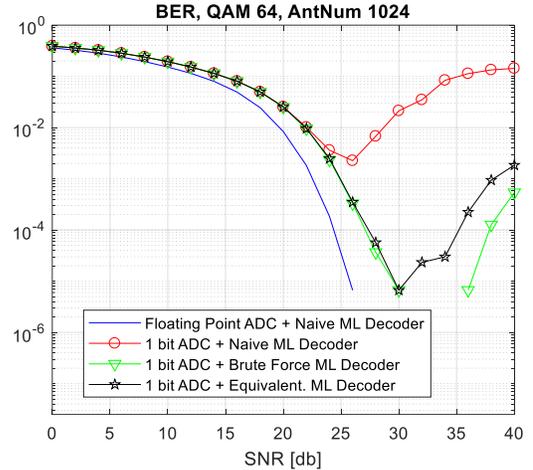

Figure 6  The BER as function of input cumulative SNR for 1024 antennas MIMO with 1 bit ADCs.

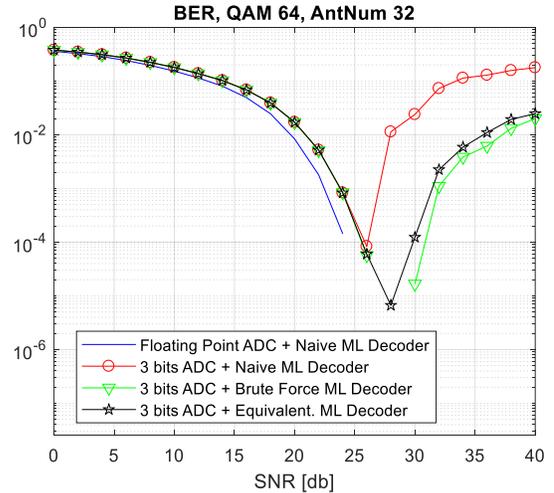

Figure 7  The BER as function of input cumulative SNR for 32 antennas MIMO with 3 bits ADCs.

## VII. CONCLUSIONS.

We proposed a much-simplified NLD-aware near maximum likelihood receiver for massive MIMO uplink equipped with low resolution ADCs, with a performance similar to that of a brute force ML receiver based on maximizing explicit likelihood functions. It demonstrates the efficacy of the new equivalent model of finite-resolution ADC detailed in [23].